\documentclass[aps,twocolumn,ams,prc,tightenlines,superscriptaddress,showpacs,floatfix,nofootinbib]{revtex4-1}

\usepackage{epsfig,bm}
\usepackage{color}
\usepackage[normalem]{ulem}  

\renewcommand\sout{\bgroup \color{red} \ULdepth=-.5ex \ULset}

\begin{document}



\title{QCD Sum Rule for Open Strange Meson $K_1^\pm$ in Nuclear Matter}


\author{Taesoo Song}%
\email{song@fias.uni-frankfurt.de}
\affiliation{Institut f\"{u}r Theoretische Physik, Universit\"{a}t Gie\ss en, Germany}
\affiliation{Frankfurt Institute for Advanced Studies, Johann Wolfgang Goethe Universit\"{a}t, Frankfurt am Main, Germany}
\author{Tetsuo Hatsuda}%
\email{thatsuda@riken.jp}

\affiliation{Interdisciplinary Theoretical and Mathematical Sciences Program (iTHEMS), RIKEN, Wako 351-0198, Japan}
\affiliation{Nishina Center, RIKEN, Wako 351-0198, Japan}

\author{Su Houng Lee}%
\email{suhoung@yonsei.ac.kr}

\affiliation{Department of Physics and Institute of Physics and Applied Physics, Yonsei
University, Seoul 03722, Korea}


\begin{abstract}
The properties of open strange meson $K_1^\pm$ in nuclear matter are
estimated in the QCD sum rule approach.
We obtain a relation between the in-medium mass and width of $K_1^-$ ($K_1^+$) in nuclear matter, and show that
the upper limit of the mass shift is as large as -249 (-35) MeV. The spectral modification
 of the $K_1$ meson is possible to be  probed by using kaon beams at J-PARC.
 Such measurement  together with that of $K^*$ will shed light on how chiral symmetry is partially  restored in nuclear matter.
\end{abstract}

 \keywords{}

\maketitle


\section{Introduction}

In the QCD vacuum,  chiral symmetry is spontaneously broken, which leads to the non-vanishing chiral order parameters and the existence of the Nambu-Goldstone (NG) bosons. It also leads to the mass difference between the vector meson and its
axial partner \cite{Weinberg:1967kj,Shifman:1978bx}. This broken symmetry is expected to have been restored in the early universe, when the
temperature was very high. Furthermore, it was noted that the chiral symmetry is partially restored
even in normal nuclear density so that by exciting mesons inside the nucleus one could study the precursor phenomena of chiral symmetry restoration \cite{Hatsuda:1985eb,Brown:1991kk,Hatsuda:1991ez,Klingl:1997kf}.
Furthermore, the enhanced repulsion of the s-wave
isovector pion-nucleus interaction observed in the deeply bound pionic atoms \cite{Suzuki:2002ae} was shown to be a direct consequence of the reduction of the in-medium quark condensate in nuclear medium \cite{Kolomeitsev:2002gc}.

According to the in-medium QCD sum rules developed in \cite{Hatsuda:1991ez},
  in-medium change of the  four-quark condensate (the strange quark condensate)
   is responsible for  the spectral change of  the $\rho, \omega$ mesons (the $\phi$ meson).
A number of  experiments have since then
carried out worldwide \cite{Hayano:2008vn}. The KEK-PS experiments observed the invariant mass spectra of $e^+~e^-$ pairs from the nuclear targets and found excess signals at the lower end of the $\omega$ resonance peak that could be  explained by the vector meson  mass decrease of 9 $\%$ at the normal nuclear density~\cite{Naruki:2005kd}. The KEK-PS E325 collaboration reported evidence that the mass of the $\phi$ meson decreased by 3.4 $\%$
at normal nuclear density \cite{Muto:2005za}.  Further measurements of dileptons from the in-medium $\phi$-meson  are planned at J-PARC E16 experiment   \cite{Ichikawa:2018woh}.
Dilepton spectrum has the advantage of not suffering from strong interaction with the medium as the signal emerges from inside the nucleus but has the disadvantage of being low in the yield. Reactions involving hadronic final states have the opposite features.
For example, CBELSA/TAPS Collaboration observed that the $\omega$ meson decreased by about 60 MeV
 at an estimated average nuclear density of 0.6 $\rho_0$ through the reaction $\gamma+A \rightarrow \omega +X \rightarrow \pi^0 \gamma +X'$ \cite{Trnka:2005ey}.
 However,  the signal could have been contaminated by final state interactions.
Furthermore, $\omega$ and $f_1$ are chiral partners only in the limit where disconnected quark diagrams are neglected \cite{Gubler:2016djf} and the QCD sum rule approach was found to have a large contribution from the scattering term \cite{DuttMazumder:2000ys,Thomas:2005dc,Leupold:2009kz},  so that it is not clear if the sum rule leads to a decreasing mass.
Alternative approach by  the CBELSA/TAPS collaboration is to extract meson-nucleus optical  potentials
 from near-threshold meson productions in  photo- and hadro- reactions off nuclei as well as in  heavy-ion reactions
  by focussing on  mesons with small width in the vacuum ($K, \eta, \eta', \omega, \phi$)
 \cite{Metag:2017yuh}. The  momentum distribution of mesons,  excitation functions and the  transparency
ratios are the key experimental observables.

Motivated by recent experimental progress, one of us (SHL) have recently estimated the spectral shift of the $f_1$ meson, which is a chiral partner of $\omega$  in the limit where the  disconnected diagrams are neglected, in the QCD sum rule approach~\cite{Gubler:2016djf}.
Experimentally,   the $f_1(1285)$ has been successfully identified by the CLAS collaboration in photoproduction
from a proton target with a small width of $18 \pm1.4$ MeV \cite{Dickson:2016gwc},  so that
 performing similar experiments on a nuclear target and comparing the result with that from the  $\omega$  would be extremely useful for
  partial restoration of chiral symmetry in nuclei as suggested in Ref.~\cite{Gubler:2016djf}.
 In fact,  the individual meson masses could behave differently depending on whether the hadron  is in nuclear matter or at finite temperature,
 while the mass difference between chiral partners will only depend on the chiral order parameter and be universal~\cite{Lee:2013es}.

 The difference between the vector and axial-vector correlation functions in the open strange channel is also an order parameter of chiral symmetry~\cite{Lee:2013es}.  This implies that their spectral densities will become degenerate if chiral symmetry is restored.  In the vacuum, the low-lying modes that couple to the vector current are  $K^*(892)$  and $K^*(1410)$ while for the axial vector current they are  $K_1(1270)$ and  $K_1(1400)$. There is a subtlety in the nature of the two $K_1$ states: They are assumed to be a mixture of the $^3P_1$ and $^1P_1$ quark-antiquark pair in the quark model \cite{Suzuki:1993yc}. However, if chiral symmetry is partially restored, the spectral density will tend to become degenerate so that the lowest distinctive poles in the respective current will approach each other. Therefore, in this work, we investigate the spectral modification of open strange meson $K_1$ through the axial-vector current in nuclear matter using QCD sum rules.

Measuring the  open strange meson in the vector channel, namely the $K^{*+}$  through the  decay  $K^{*+} \rightarrow K^+ + \gamma$ was suggested as a promising signal to measure the spectral change of the vector meson in Ref.~\cite{Hatsuda:1997ev}.
Both the $K_1(1270) $ and the $K^{*}(892)$ have widths smaller than their non strange counter parts, namely 90 MeV and 47 MeV, respectively, compared to more than 250 MeV and 150 MeV for the $a_1$ and the $\rho$. At the same time, they are also chiral partners so that their mass difference is sensitive to the chiral order parameter.

We note that  $K_1^+$ and $K_1^-$ become non-degenerate in nuclear medium due to the  presence of nucleons which break charge conjugation  invariance in the medium.
 There are two approaches to treat such situation in QCD sum rules.
    One is to project out the  polarization function into definite charge conjugation states \cite{Jido:1996ia,Suzuki:2015est}.  The other is to extract the ground state of each charge state from the polarization functions \cite{Kondo:2005ur}.  In the present paper, we take the latter method in which
     some parameters for $K_1^-$ are mixed into the sum rule for $K_1^+$ and vice versa.

In section II, we first discuss the QCD sum rules for $K_1$ meson in the vacuum.
 In section III, effects of nuclear matter  are taken into account  in the QCD sum rule through the
local operators with spin. In section III, the maximum mass shift  of $K_1^\pm$ meson is estimated
by the pole plus continuum approximation of the  spectral function. Also,  by considering the  modification of the width and the mass shift, we
 obtain a relation  between the in-medium change of these two quantities.  Summary and discussion  are given in section IV.

\section{$K_1$ meson in vacuum}
The time-ordered current correlation function of the $K_1$ current is given by

\begin{eqnarray}
\Pi_{\mu \nu}(q)= i\int d^4 x e^{iq\cdot x}\langle 0|T[\bar{u}\gamma_\mu \gamma_5
s(x), \bar{s}\gamma_\nu \gamma_5 u(0)]|0\rangle\nonumber\\
= -i\int d^4 x e^{iq\cdot x}\langle 0|{\rm Tr}[\gamma_\mu \gamma_5 iS_s(x) \gamma_\nu \gamma_5 iS_u
(-x)]|0\rangle, \nonumber\\
\label{correlator}
\end{eqnarray}
where $iS_s$ and $iS_u$ represent u-quark and s-quark propagators respectively. Including
nonperturbative effect, these propagators are expanded as \cite{Furnstahl:1992pi}:
\begin{eqnarray}
iS^{a b}(x) = i\frac{\not{x}}{2\pi^2
x^4}\delta^{ab}-\frac{m}{4\pi^2x^2}\delta^{ab}+i\frac{m^2
\not{x}}{8\pi^2 x^2}\delta^{ab}+...\nonumber\\
+ \chi^a(x) \chi^b(0) -i\frac{g}{32
\pi^2}\frac{\not{x}\sigma_{\alpha \beta}+\sigma_{\alpha
\beta}\not{x}}{x^2}F^{\alpha \beta}_A (0) t_A^{ab}+., \label{propagator}
\end{eqnarray}
where $a$, $b$ are color indices, and $\alpha$, $\beta$ Lorentz indices. The first line is the expansion of the perturbative part with respect to quark mass $m$, and the second line encodes the nonperturbative part. $\chi^a$
is the background field of the quark, and $F^{\alpha \beta}_A$ that of the
gluon. Because the masses of u and d quarks are small compared to the typical QCD scale, we consider only strange quark mass to be finite in this study.

The current correlation function in the vacuum is composed of two independent functions  $\Pi_1$ and $\Pi_2$ as follows:
\begin{equation}
\Pi_{\mu \nu}(q)=-g_{\mu \nu}\Pi_1(q^2)+ q_\mu q_\nu \Pi_2(q^2).
\end{equation}
If the current is conserved, the two functions are related by $\Pi_1=q^2\Pi_2$. However, the axial current of $K_1$ is not conserved and $\Pi_2$ has contributions from pseudoscalar mesons. In principle, we may carry out QCD sum rule with either $\Pi_1$ or  $\Pi_2$.

Let us first consider $\Pi_1$.
Substituting Eq. (\ref{propagator}) into Eq. (\ref{correlator}), the operator
 product expansion (OPE) of the current correlation function is obtained up to dimension 6 as
\begin{eqnarray}
\Pi_1(q^2)=B_0 Q^2\ln \frac{Q^2}{\mu^{2} }+ B_2\ln \frac{Q^2}{\mu^2}
-\frac{B_4}{Q^2}-\frac{B_6}{Q^4},
\label{OPE}
\end{eqnarray}
and
\begin{eqnarray}
B_0 &=& \frac{1}{4\pi^2}\bigg(1+\frac{\alpha_s}{\pi}\bigg)\nonumber\\
B_2 &=& \frac{3m_s^2}{8 \pi^2}\nonumber\\
B_4 &=& -m_s \langle\bar{u}u\rangle_0 +\frac{1}{12}\langle \frac{\alpha_s}{\pi} G^2\rangle_0 \nonumber\\
B_6
 &= & \frac{2 \pi \alpha_s}{9 } \bigg( \langle (\bar{s} \gamma_\mu \lambda^a s +\bar{u} \gamma_\mu \lambda^a u  )( \sum_q \bar{q} \gamma_\mu \lambda^aq )\rangle_0 \bigg)\nonumber\\
 && + 2 \pi \alpha_s \bigg( \langle (\bar{u} \gamma_\mu \lambda^a s )( (\bar{s} \gamma_\mu \lambda^a u ) \rangle_0
 \bigg)
 \nonumber \\
&=& \frac{32\pi \alpha_s}{81}\bigg(\langle\bar{u}u\rangle_0^2+\langle\bar{s}s\rangle_0^2\bigg)  +
\frac{32\pi \alpha_s}{9}\bigg(\langle\bar{u}u\rangle_0 \langle\bar{s}s\rangle_0\bigg), \nonumber
\end{eqnarray}
where $Q^2 \equiv -q^2$, $\mu=1$ GeV is the renormalization scale,
 $n$ in $B_n$ indicates the canonical dimension of the operator, and $\langle {\cal O} \rangle_0$ denotes the condensate of operator $\cal{O}$ in the vacuum.
 We take the recently updated parameters as follows: $\alpha_s=0.5$, $m_q=4.26$ MeV, $m_s=117$ MeV, where both quark masses are scaled to $\mu=1$ GeV from the values at $\mu=2$ GeV given by lattice calculations in 2+1 flavors as reported in Particle Data Group~\cite{Tanabashi:2018oca}, $\langle\bar{u}u\rangle_0=(-0.262~{\rm GeV})^3$ from the Gell-Mann Oakes Renner relation $2m_q\langle\bar{u}u\rangle_0=m_\pi^2 f_\pi^2$ with $m_\pi$ and $f_\pi$ being the mass and decay constant of the pion respectively, $\langle\bar{s}s\rangle_0=0.8\langle\bar{u}u\rangle_0$, and $\langle \alpha_s /\pi G^2\rangle_0=0.012~ {\rm GeV}^4$.
\footnote{A recent work finds a larger gluon condensate from the analysis of the $e^+e^-$ annihilation data in the charm-quark region \cite{Dominguez:2014pga}.
  Considering the fact that the higher order $\alpha_s$  corrections and the value of the gluon condensate are correlated\cite{Novikov:1984rf,Lee:1989qj},
  we choose, in this paper,  a parameter set  (Table II of \cite{Gubler:2016djf}) that reproduces the masses and decay constants of the light quark system consistently in the leading order.}
For the four quark operators of dimension 6, a factorization ansatz is adopted \cite{Shifman:1978bx,Hatsuda:1991ez}.

It should be noted that for the corresponding vector correlation function obtained with the current $J^{K^*}_\mu=\bar{u} \gamma_\mu s$, the OPE up to this order will be similar with $B_0^V=B_0$, $B_2^V=B_2$ and
\begin{eqnarray}
B_4^V &=& +m_s \langle\bar{u}u\rangle_0 +\frac{1}{12}\langle \frac{\alpha_s}{\pi} G^2\rangle_0 \nonumber\\
B_6^V
 &= & \frac{2 \pi \alpha_s}{9 } \bigg( \langle (\bar{s} \gamma_\mu \lambda^a s +\bar{u} \gamma_\mu  \lambda^a u)( \sum_q \bar{q} \gamma_\mu \lambda^aq) \rangle_0\bigg) \nonumber\\
 && + 2 \pi \alpha_s \bigg( \langle (\bar{u} \gamma_\mu \gamma^5 \lambda^a s )( (\bar{s} \gamma_\mu \gamma^5 \lambda^a u ) \rangle_0
 \bigg)
 \nonumber \\
&=& \frac{32\pi \alpha_s}{81}\bigg(\langle\bar{u}u\rangle_0^2+\langle\bar{s}s\rangle_0^2\bigg)  -
\frac{32\pi \alpha_s}{9}\bigg(\langle\bar{u}u\rangle_0\langle\bar{s}s\rangle_0\bigg). \nonumber
\end{eqnarray}
The difference between the axial and the vector correlation functions is
proportional to chiral symmetry breaking operators responsible for the mass difference between chiral partners.
 Specifically, at dimension 4, the difference is proportional to $m_s \langle \bar{q} q \rangle$, while at dimension 6,  it is $\langle \bar{s} s \rangle \langle \bar{q}q \rangle$.   Both operators are proportional to $\langle \bar{q} q \rangle$, but is dominated by the dimension 6 operator. It is interesting to note that the larger mass difference between $  m_{a_1} -m_\rho \simeq 1260-770= 490 $ MeV compared to the corresponding mass difference in the open strange sector $m_{K_1}-m_{K^*} \simeq 1270-892=378$ MeV seems to be related to the difference in the four quark condensate $\langle \bar{q}q \rangle^2$ to  $\langle \bar{s} s \rangle \langle \bar{q}q \rangle$ in their respective sum rules.  It should be also noted that the difference in the open charm sector is dominated by the dimension 4 operators because the charm quark mass amplifies the contribution from the light quark condensate as was noted in Ref. \cite{Hilger:2011cq,Buchheim:2014rpa}.

As for the imaginary part of the correlation function, we use the phenomenological spectral function
\begin{eqnarray}
\frac{1}{\pi}{\rm Im}\Pi_1 (q^2)&=&  \frac{m_{K_1}^4}{g_{K_1}^2}\delta(q^2-m_{K_1}^2)\nonumber\\
&&+(B_0 q^2-B_2) \theta(q^2-s_0),
\label{imaginary-vacuum}
\end{eqnarray}
where the first term on the right hand side represents the ground state and the second term the sum of all excited states, which is approximated by the continuum part starting from a threshold value $s_0$. The factor multiplied to the step function is obtained by the perturbative part of Eq.~(\ref{OPE}), as shown in Appendix A.
The imaginary part and the real part of the correlation function are related to each other through the dispersion relation:
\begin{eqnarray}
\frac{1}{\pi}\int \frac{{\rm Im} \Pi_1(s)}{s+Q^2}ds={\rm Re} \Pi_1(q^2).
\label{dispersion}
\end{eqnarray}

In order to improve our approximations from both sides, namely the calculations of the real part up to dimension 6 and the step function for excited states and continuum, we take the Borel transformation defined as
\begin{eqnarray}
\hat{B}\equiv \lim_{Q^2, n \rightarrow \infty~ Q^2/n \rightarrow M^2}\frac{1}{(n-1)!}(Q^2)^n\bigg(-\frac{d}{dQ^2}\bigg)^n, \nonumber
\end{eqnarray}
where $M$ is called the Borel mass.  For our purpose, we use
\begin{eqnarray}
&&\hat{B}(Q^2)^{-k} = \frac{1}{(k-1)!}\frac{1}{(M^2)^k},\label{power}\\
&&\hat{B}(Q^2)^{k}\ln Q^2 = -k!(-M^2)^k,\\
&&\hat{B}(Q^2+M_B^2)^{-k} = \frac{1}{(k-1)!}\frac{1}{(M^2)^k}e^{-M_B^2/M^2}.
\label{expo}
\end{eqnarray}

Taking the Borel transformation has two advantages. First, as shown in Eq.~(\ref{expo}), it introduces an exponential function in the left hand side of Eq.~(\ref{dispersion}), which enhances the ground state but suppresses the continuum part.
Second, the contribution from high-dimension operators in the real part of the polarization function is suppressed by an additional $1/(n-1)!$ factor.

Substituting the OPE and the phenomenological ansatz into Eq. (\ref{dispersion}), we obtain the following equation after the  Borel transformation.
\begin{eqnarray}
&&\frac{m_{K_1}^4}{g_{K_1}^2} e^{-m_{K_1}^2/M^2}= B_0 M^4 \bigg\{1-\bigg(1+\frac{s_0}{M^2}\bigg)e^{-s_0/M^2}\bigg\} \nonumber\\
&&~~~~~~~~ - B_2 M^2 \bigg(1-e^{-s_0/M^2}\bigg)-B_4-\frac{B_6}{M^2},
\label{dispersion-2}
\end{eqnarray}
where the continuum part was moved to the right hand side.
Differentiating Eq.~(\ref{dispersion-2}) with respect to $1/M^2$, and dividing it by Eq. (\ref{dispersion-2}), $K_1$ mass is expressed as
\begin{eqnarray}
m_{K_1}^2=M^2\frac{2 B_0 E_2   -B_2 E_1/M^2 +B_6/M^6}
{B_0 E_1 -B_2 E_0/M^2
-B_4/M^4 -B_6/M^6 }, \nonumber \\
\label{borel-curve}
\end{eqnarray}
where
\begin{eqnarray}
E_0 &=& 1-e^{-s_0/M^2},\nonumber\\
E_1 &=& 1-\bigg(1+\frac{s_0}{M^2}\bigg)e^{-s_0/M^2},\nonumber\\
E_2 &=& 1-\bigg(1+\frac{s_0}{M^2}+\frac{s_0^2}{2M^4}\bigg)e^{-s_0/M^2}.\nonumber
\end{eqnarray}

Eq. (\ref{borel-curve}) is the Borel sum rule for  the $K_1$ mass.
In principle, the physical mass should be independent of $M^2$. However, as mentioned above, the real part of the correlation function is truncated at dimension 6, and the excited states and continuum in the spectral function is simplified into a step function. As a result Eq.~(\ref{borel-curve}) depends on $M^2$.
Then, one introduces the so-called Borel window in $M^2$ where the resultant $K_1$ mass is reliable.
The smallest $M^2$ of the Borel window, $M^2_{\rm min}$, is determined from the condition that
the contribution from the power corrections does not exceed 15 \% of the perturbative part:

\begin{eqnarray}
\bigg| \frac{B_4 + B_6/M^2}{B_0 M^4 -B_2 M^2} \bigg| < 0.15.
\label{power}
\end{eqnarray}

The largest reliable $M^2$, $M^2_{\rm max}$, is determined from the condition that
the contribution from the continuum does not exceed 70 \%:
\begin{eqnarray}
\bigg| \frac{B_0 M^2 (1-E_1)-B_2(1-E_0)}{B_0 M^2-B_2 }\bigg| < 0.7.
\label{continuum}
\end{eqnarray}
We note that the maximum percentage is taken to be larger than in Eq.~(\ref{power}), which is similar to the continuum contribution for the p-wave states using the $\Pi_1$ sum rule~\cite{Reinders:1981ww}.
If $M_{\rm min}^2$  becomes too small the large power correction spoils the stability, while if the $M_{\rm max}^2$ becomes too small the sensitivity of the continuum threshold is lost and one needs a large change in the continuum threshold.

Applying Eqs.~(\ref{power}) and (\ref{continuum}), the Borel window is given by $1.06 \leq M^2 \leq 2.17~{\rm GeV^2}$.
The continuum threshold $s_0$ is chosen such that the extremum of the Borel curve is close to the physical mass of $K_1$ ground state within the Borel window.
Physically it should be close to the mass of the first excited state.

\begin{figure}
\centerline{\includegraphics[width=10 cm]{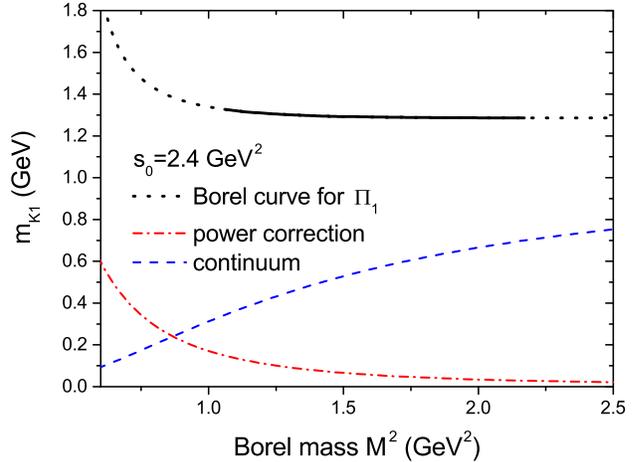}}
\caption{ The Borel curve for $K_1$ mass from the $\Pi_1$ sum rule together with the fractional contributions from the power corrections and the continuum. The Borel window given in Eqs.~(\ref{power}) and (\ref{continuum}) is shown by the black solid line. The unit for the mass is in GeV while for the fractional contributions it is unitless. }
\label{k1-vacuum}
\end{figure}

Figure~\ref{k1-vacuum}  shows the Borel curve for the mass of $K_1$ at $s_0=2.4~{\rm GeV^2}$ together with the fractional contributions from the power corrections and the continuum. The Borel window which satisfies Eqs.~(\ref{power}) and (\ref{continuum}) is shown by the black solid line. We find
 the minimum value of the Borel curve is consistent with the mass of  $K_1(1270)$.
 The overlap strength of the current with the ground state  $F_{K_1}\equiv m_{K_1}^2/g_{K_1}^2$ is about 0.048 ${\rm GeV^2}$ in this window.
We note that $\sqrt{s_0}=1.55~{\rm GeV}$ is close to the mass of $K_1(1400)$.

We can also construct a  QCD sum rule with $\Pi_2$. The real part is then given by
\begin{eqnarray}
\Pi_2(q^2)=-B_0 \ln \frac{Q^2}{\mu^{2} }- \frac{\tilde{B}_2}{Q^2}
+\frac{\tilde{B}_4}{Q^4}+\frac{B_6}{Q^6},
\end{eqnarray}
where
\begin{eqnarray}
\tilde{B}_2 &=& \frac{3m_s^2}{4 \pi^2},\nonumber\\
\tilde{B}_4  &=& m_s \langle\bar{s}s\rangle_0 +\frac{1}{12}\langle \frac{\alpha_s}{\pi} G^2\rangle_0,\nonumber
\end{eqnarray}
and $B_0$ and $B_6$ are same as in $\Pi_1$, with $\tilde{B}_4=B_4$ in the limit $m_s \rightarrow 0$.  However, the imaginary part will then have contribution from the pseudoscalar meson and will not be useful for our purpose as one would need  additional input on the kaon properties in medium to  use the corresponding sum rule to study the properties of the $K_1$ meson in medium.
Therefore, we use the Borel curve from $\Pi_1$ in this study to investigate the properties of $K_1$ meson in nuclear matter.

\section{$K_1^\pm$ meson in nuclear medium}

In nuclear matter, there are two modifications in the real part of the  correlation function. One is the change in the values of the  condensates, and the other is the appearance of operators with spin.
We take into account only twist-2 terms which are dominant in OPE. (Here twist is the dimension of operator subtracted by spin.)  Higher twist terms  have been estimated before and are known to be less important \cite{Friman:1999wu}.
The degeneracy of  $K_1^-$ and $K_1^+$ in the vacuum does not hold   in  nuclear matter due to charge symmetry breaking,
which leads the odd dimensional terms in the OPE to contribute with opposite sign for the two charged states.

\subsection{In-medium OPE}

In the limit ${\bf q} \rightarrow 0$, the OPE in nuclear matter can be written as
\begin{eqnarray}
\Pi_1(q^2)=\Pi^e(q^2)+q_0 \Pi^o(q^2),\label{pi1eo-m}
\end{eqnarray}
where
\begin{eqnarray}
\Pi^e(q^2)&=& B_0 Q^2\ln \frac{Q^2}{\mu^{2} }+ B_2\ln \frac{Q^2}{\mu^2}
-\frac{B_4^*}{Q^2}-\frac{B_6^*}{Q^4},\nonumber\\
\Pi^o(q^2) &=& \frac{1}{3Q^2}(A_1^u-A_1^s)\rho
-\frac{2m_N^2}{3Q^4}(A_3^u-A_3^s)\rho, \nonumber \\
&& \label{pieo}
\end{eqnarray}
with
\begin{eqnarray}
&&B_4^*=-m_s \langle\bar{u}u \rangle_\rho
+\frac{1}{12}\langle \frac{\alpha_s}{\pi} G^2\rangle_\rho, \nonumber\\
&&~~~~~~~ +\frac{m_N}{2}(A_2^u +A_2^s)\rho \nonumber\\
&&B_6^*=\frac{32\pi \alpha_s}{9}\bigg\{
\langle\bar{u}u \rangle_\rho \langle\bar{s}s\rangle_\rho +\frac{\langle\bar{u}u\rangle_\rho^2+\langle\bar{s}s\rangle_\rho^2}{9}\bigg\}\nonumber\\
&&~~~~~~~ -\frac{5}{6}m_N^3(A_4^u +A_4^s)\rho.
\label{ope-medium}
\end{eqnarray}

$\Pi^{e(o)}$ denotes even(odd) dimensional terms of correlation function. $\langle {\cal O}\rangle_\rho$ is the condensate of operator ${\cal O}$ in nuclear matter, where we use the linear density approximation: $\langle {\cal O}\rangle_\rho=\langle {\cal O}\rangle_0+\langle {\cal O}\rangle_N  \rho$ with $\rho$ being the baryon density.  $\langle\bar{u}u\rangle_N$, $\langle\bar{s}s\rangle_N$, and $\langle G^2\rangle_N$ are the nucleon matrix elements taken from Refs.~\cite{Hatsuda:1991ez,Gubler:2016djf}:
\begin{eqnarray}
m_q\langle \bar{u}u+\bar{d}d\rangle_N = 45~{\rm MeV},~m_s\langle \bar{s}s\rangle_N = 35~{\rm MeV},\nonumber\\
\langle \alpha/\pi G^2\rangle_N=-8/9m_N.~~~~~~~~~~~~
\end{eqnarray}
$m_N$ is the nucleon mass, and $A_n^q(\mu^2)=2\int_0^1
x^{n-1}\{q(x,\mu^2)+(-1)^n \bar{q}(x,\mu^2)\}dx$, where $q(x,\mu^2)$ and $\bar{q}(x,\mu^2)$ are, respectively,
quark and antiquark distribution functions in the nucleon at scale $\mu^2$, and are defined through the twist-two operators
\begin{eqnarray}
\langle {\cal ST}(\bar{q}\gamma_{\mu_1} D_{\mu_2} ... D_{\mu_n}q(\mu^2))\rangle_N  \nonumber \\
=(-i)^{n-1}A_n^q(\mu^2)\frac{T_{\mu_1 ...\mu_n}}{2m_N}.
\end{eqnarray}
Here, ${\cal ST}$ means `symmetric and traceless',  and expressed on the right side  with the tensor $T_{\mu_1 \cdots\mu_n}$. For our case,
\begin{eqnarray}
T_{\mu \nu}&=& p_\mu p_\nu -\frac{p^2}{4}g_{\mu \nu},\nonumber\\
T_{\mu \nu \rho}&=& p_\mu p_\nu p_\rho -\frac{p^2}{6}(g_{\mu \nu}p_\rho +
+g_{\mu \rho}p_\nu +g_{\nu \rho}p_\mu), \nonumber\\
T_{\mu \nu \rho \lambda}&=& p_\mu p_\nu p_\rho p_\lambda -\frac{p^2}{8}(g_{\mu \nu}p_\lambda p_\rho +
g_{\mu \lambda}p_\nu p_\rho \nonumber\\
&&+g_{\mu \rho}p_\nu p_\lambda +g_{\nu \lambda}p_\mu p_\rho +g_{\nu \rho}p_\mu p_\lambda +g_{\lambda \rho}p_\mu p_\nu) \nonumber\\
&&+\frac{p^4}{48}(g_{\mu \nu}g_{\lambda \rho}+g_{\mu \lambda}g_{\nu \rho}+g_{\mu \rho}g_{\nu \lambda}),
\end{eqnarray}
where $p_\mu$ is nucleon four momentum.  We calculate $A_n^q(\mu^2)$
  using the MSTW parton distribution function~\cite{Martin:2009iq} at the scale $\mu^2=1 ~{\rm GeV^2}$, which is same as our renormalization scale of 1 GeV:
   \begin{eqnarray}
A_1^u&=&3.0,~~~A_2^u=0.62,~~~A_3^u=0.15,~~~A_4^u=0.0637, \nonumber\\
A_1^s&=&0.0,~~~A_2^s=0.048,~~~A_3^s=0.00085,~~~A_4^s=0.0011. \nonumber
\end{eqnarray}

\subsection{Phenomenological side}

$\Pi^{e}$  and $\Pi^{o}$ appearing in Eq.~(\ref{pi1eo-m}) are respectively  even and odd under charge conjugation.  Therefore,
 the charge even and odd states will become non-degenerate  in the medium,  so that we have to introduce separate physical states for the positive and negative charge states.
Then the  imaginary part in Eq.~(\ref{pi1eo-m})  can  be written as
\begin{eqnarray}
\frac{1}{\pi}{\rm Im}\Pi_1 (q^2)=  \frac{m_{K_1^-}^3}{2g_{K_1^-}^2}\delta(q_0-m_{K_1^{-}})+ \frac{m_{K_1^+}^3}{2g_{K_1^+}^2}\delta(q_0+m_{K_1^{+}})
\nonumber\\
+ (B_0 q^2-B_2) \bigg\{ \theta\bigg(q_0-\sqrt{s_0^-}\bigg)+\theta\bigg(-q_0-\sqrt{s_0^+}\bigg) \bigg\}. \nonumber \\
\label{imaginary-1}
\end{eqnarray}
 in the limit ${\bf q} \rightarrow 0$.
Using the definition of Eq.~(\ref{pieo}), we can then extract  $\Pi^e$ and $\Pi^o$ separately.   Since we are interested in the $K_1^{\pm}$ state separately, we
consider the following combination of the polarization function;
\begin{eqnarray}
\frac{2}{\pi}{\rm Im}(\Pi^e +m_{K_1^+} \Pi^o)= \frac{m_{K_1^-}^4}{ g_{K_1^-}^2}
\bigg(1+\frac{m_{K_1^+}}{m_{K_1^-}}\bigg)\delta(q^2-m_{K_1^-}^2)&&\nonumber\\
+(B_0 q^2-B_2)\bigg\{ (1+\frac{m_{K_1^+}}{q_0})\theta(q^2-s_0^-)~~~~~&&\nonumber\\
+(1-\frac{m_{K_1^+}}{q_0})\theta(q^2-s_0^+)\bigg\},~~&&\nonumber\\
\frac{2}{\pi}{\rm Im}(\Pi^e -m_{K_1^-} \Pi^o)= \frac{m_{K_1^+}^4}{ g_{K_1^+}^2}
\bigg(1+\frac{m_{K_1^-}}{m_{K_1^+}}\bigg)\delta(q^2-m_{K_1^+}^2)\nonumber&&\\
+(B_0 q^2-B_2)\bigg\{ (1+\frac{m_{K_1^-}}{q_0})\theta(q^2-s_0^+)~~~~~&&\nonumber\\
+(1-\frac{m_{K_1^-}}{q_0})\theta(q^2-s_0^-)\bigg\}.~&&\nonumber\\
\label{imaginary-medium}
\end{eqnarray}

The first equation has the resonance of $K_1^-$ and the continuum of $K_1^-$ and $K_1^+$. On the other hand, the second equation has the resonance of $K_1^+$ and the continuum of $K_1^-$ and $K_1^+$.  The detailed derivation of the imaginary part and Eq.~(\ref{imaginary-medium}) is given in Appendix B.
We note that the two equations in  Eq.~(\ref{imaginary-medium}) reduce to Eq. (\ref{imaginary-vacuum}), when $m_{K_1^-}=m_{K_1^+}$, $g_{K_1^-}=g_{K_1^+}$, and $s_0^-=s_0^+$.
We can see on the left hand side of Eq.~(\ref{imaginary-medium}) that the odd dimensional terms contribute to $K_1^-$ with a positive sign and to $K_1^+$ with a negative sign. It brings about the splitting between $K_1^-$ and $K_1^+$ in nuclear matter.

\subsection{In-medium QCD sum rules}

The dispersion relation for each combination
\begin{eqnarray}
\frac{1}{\pi}\int \frac{{\rm Im} ~(\Pi^e \pm m_{K_1^\pm}\Pi^o)ds}{s+Q^2}={\rm Re} ~ (\Pi^e \pm m_{K_1^\pm} \Pi^o),\nonumber\\
\end{eqnarray}
reduces to the following equations after the Borel transformation;

\begin{eqnarray}
&&\frac{F_\pm m_\pm^2}{2M^2}\bigg(1+\frac{m_\mp}{m_\pm}\bigg)e^{-m_\pm^2/M^2}\nonumber\\
&&+\frac{1}{2M^2}\int_{s_0^\pm}^\infty \bigg(1+\frac{m_\mp}{\sqrt{s}}\bigg)(B_0s-B_2)e^{-s/M^2}ds\nonumber\\
&&+\frac{1}{2M^2}\int_{s_0^\mp}^\infty \bigg(1-\frac{m_\mp}{\sqrt{s}}\bigg)(B_0s-B_2)e^{-s/M^2}ds
\nonumber\\
&&=B_0 M^2-B_2-\frac{1}{M^2}\bigg\{B_4^*\pm \frac{m_\mp}{3}(A_1^u-A_1^s)\rho\bigg\}\nonumber\\
&&~~~-\frac{1}{M^4}\bigg\{B_6^*\mp \frac{2m_\mp m_N^2}{3}(A_3^u-A_3^s)\rho \bigg\},
\label{Borel-medium}
\end{eqnarray}
where $m_{K_1^\pm}$ and $m_\pm^2/g_{K_1^\pm}^2$ are respectively abbreviated to $m_\pm$ and $F_\pm$.

\subsubsection{In-medium mass}

For small nuclear density,  the overlap strength, mass, and continuum threshold  may be approximated as

\begin{eqnarray}
F_\pm &=& F_{K_1}+F'_\pm \rho, \nonumber\\
m_\pm &=& m_{K_1}+m'_\pm  \rho, \nonumber\\
s_0^\pm &=& s_0+{s'}_0^\pm \rho.
\end{eqnarray}

Keeping only terms linearly proportional to $\rho$, Eq. (\ref{Borel-medium}) reduces to
\begin{eqnarray}
F(M^2)F'_\pm +M(M^2)m'_\pm +S(M^2){s'}_0^\pm =C_\pm(M^2),\nonumber\\
\label{eq:SR-m}
\end{eqnarray}

where
\begin{eqnarray}
&&F(M^2)= -m_{K_1}^2e^{-m_{K_1}^2/M^2},\nonumber\\
&&M(M^2)= F_{K_1}m_{K_1}\bigg(-\frac{3}{2}+\frac{2m_{K_1}^2}{M^2}\bigg)e^{-m_{K_1}^2/M^2},\nonumber\\
&&S(M^2)= \frac{1}{2}\bigg(1+\frac{m_{K_1}}{\sqrt{s_0}}\bigg)(B_0 s_0-B_2)e^{-s_0/M^2},\nonumber\\
&&C_\pm(M^2)=-m_s \langle\bar{u}u\rangle_N
+\frac{\alpha_s}{12\pi}\langle G^2\rangle_N\nonumber\\
&&~~~+\frac{m_N}{2}(A_2^u +A_2^s)\pm \frac{m_{K_1}}{3}(A_1^u-A_1^s)\nonumber\\
&&~~~+\frac{32\pi \alpha_s}{9 M^2}\bigg\{
\langle\bar{u}u\rangle_N \langle\bar{s}s\rangle_0 +\langle\bar{u}u\rangle_0 \langle\bar{s}s\rangle_N\nonumber\\
&&~~~~~~~~~~~+\frac{2}{9}(\langle\bar{u}u\rangle_N \langle\bar{u}u\rangle_0+\langle\bar{s}s\rangle_N\langle\bar{s}s\rangle_0)\bigg\}\nonumber\\
&&~~~-\frac{5m_N^3}{6 M^2}(A_4^u +A_4^s)\mp \frac{2m_{K_1} m_N^2}{3M^2}(A_3^u-A_3^s)\nonumber\\
&&~~~+m'_\mp \bigg\{\frac{F_{K_1}m_{K_1}}{2}e^{-m_{K_1}^2/M^2}\bigg\}\nonumber\\
&&~~~+{s'}_0^\mp \bigg\{\frac{1}{2}\bigg(-1+\frac{m_{K_1}}{\sqrt{s_0}}\bigg)(B_0 s_0-B_2)e^{-s_0/M^2}\bigg\}.\nonumber\\
\label{minimum1}
\end{eqnarray}

We now define $V_\pm(F'_\pm, m'_\pm, {s'}_0^\pm)$ from Eq.~(\ref{eq:SR-m}) as
\begin{eqnarray}
&&V_\pm(F'_\pm, m'_\pm, {s'}_0^\pm)\equiv
\int_{M^2_{\rm min}}^{M^2_{\rm max}} \bigg\{F(M^2)F'_\pm +M(M^2)m'_\pm \nonumber\\
&&~~~~~~~~~~~~~~~~~~+S(M^2){s'}_0^\pm - C_\pm(M^2)\bigg\}^2 dM^2,
\label{minimum2}
\end{eqnarray}
where $M^2_{\rm min}$ and $M^2_{\rm max}$ are the lower and upper limits of the Borel window, which are respectively taken to be 1.06 and 2.17 ${\rm GeV^2}$ as in the vacuum.  This will be justified in the Borel analysis discussed in the next section, where we show that the most stable Borel curve has a plateau within this Borel window and that the obtained mass shift and threshold change are consistent with those calculated in this section.
Though Eq. (\ref{minimum2}) is supposed to vanish in the ideal case, it is always positive because of the approximations taken both in the OPE side and in the  phenomenological side. Therefore, we search for $F'_\pm, m'_\pm$, and ${s'}_0^\pm $ which minimize the function $V_\pm$, that is,
\begin{eqnarray}
\frac{\partial V_\pm}{\partial F'_\pm }=\frac{\partial V_\pm}{\partial m'_\pm }=\frac{\partial V_\pm}{\partial {s'}_0^\pm}=0.
\end{eqnarray}

These conditions result in three simultaneous linear equations as follows:
\begin{widetext}
\begin{eqnarray}
F'_\pm \int dM^2 F^2(M^2) +m'_\pm \int dM^2 F(M^2)M(M^2)
+{s'}_0^\pm\int dM^2 F(M^2)S(M^2)  &=& \int dM^2 F(M^2) C_\pm(M^2), \nonumber\\
F'_\pm \int dM^2 F(M^2)M(M^2) +m'_\pm \int dM^2 M^2(M^2)
+{s'}_0^\pm \int dM^2 M(M^2)S(M^2) &=& \int dM^2 M(M^2) C_\pm(M^2), \nonumber\\
F'_\pm \int dM^2 F(M^2)S(M^2) +m'_\pm \int dM^2 M(M^2)S(M^2)
+{s'}_0^\pm \int dM^2 S^2(M^2) &=& \int dM^2 S(M^2) C_\pm(M^2) .
\label{coupled1}
\end{eqnarray}

\begin{table}
\centering
\begin{tabular}{|ccc|ccc|}
\hline \hline
 $F'_- \rho_0\ ({\rm GeV^2})$ & $m'_-\rho_0\ ({\rm GeV})$ & ${s'}_0^-\rho_0\ ({\rm GeV^2})$ & $F'_+\rho_0\ ({\rm GeV^2})$ & $m'_+\rho_0\ ({\rm GeV})$ & ${s'}_0^+\rho_0\ ({\rm GeV^2})$  \\
 \hline
-3.09$\times 10^{-2}$ & -0.249 & -1.25 & -2.72$\times 10^{-3}$ & -0.0348 & -0.234   \\
\hline \hline

\end{tabular}
\caption{  Results for $F'_\pm \rho$, $m'_\pm \rho$, and
${s'}_0^\pm \rho$ at  normal nuclear matter density
$\rho=\rho_0=0.16 ~{\rm fm^{-3}}$.
The mass shift at nuclear matter density is $\delta m_{K_1^\pm}=m'_\pm\rho$. }
\label{iteration}
\end{table}
\end{widetext}

The above equations are coupled such that  $V_+$ is a function of $m_-$ and $s_0^-$ as well as of $m_+$ and $s_0^+$, while $V_-$ is a function of $m_+$ and $s_0^+$ as well as of $m_-$ and $s_0^-$.  This is so because $C_\pm(M^2)$ are functions of $m_\mp$ and $s_0^\mp$. In other words, three simultaneous linear equations for $F'_+$, $m_+$,
and ${s'}_0^+ $ are coupled with those for $F'_-$, $m'_-$, and ${s'}_0^- $. We solve these equations  iteratively,  with
the final results of  $F_\pm {'}$, $m'_\pm $, and ${s'}_0^\pm $ shown in Table \ref{iteration}.
We find that with the obtained values, the  ratios of
 $V_\pm(F'_\pm, m'_\pm, {s'})$ in  Eq.~(\ref{minimum2}) to the individual terms,
$\int_{M^2_{\rm min}}^{M^2_{\rm max}} ( F(M^2)F'_\pm )^2 dM^2$, $\int_{M^2_{\rm min}}^{M^2_{\rm max}} ( M(M^2)m'_\pm )^2 dM^2$,  $ \int_{M^2_{\rm min}}^{M^2_{\rm max}} (S(M^2){s'}_0^\pm)^2 dM^2$, and $\int_{M^2_{\rm min}}^{M^2_{\rm max}} ( C_\pm(M^2))^2 dM^2 $
are (1.8$\times 10^{-6}$, 2.2$\times 10^{-5}$, 3.8$\times 10^{-7}$, 9.5$\times 10^{-7}$) and (1.7$\times 10^{-7}$, 5.3$\times 10^{-6}$, 1.7$\times 10^{-7}$, 7.3$\times 10^{-6}$), respectively for $\pm$ states, justifying our optimization procedure.

In these calculations, $m_{K_1}$ and $s_0$ are set to be  $\rm 1.27~GeV$ and $\rm 2.4 ~GeV^2$, respectively. As a result, the mass of $K_1^-$ decreases by 249 $\rm~MeV$
 in nuclear matter, which corresponds to 20 \% reduction of the mass from its vacuum value.
  On the other hand, the mass of $K_1^+$ decreases only by 35 $\rm~MeV$, which is about 3 \% of its vacuum mass.
Since we have neglected the in-medium width in this subsection, these numbers are the upper limits of the mass shift.
 Under such condition, the result indicates that $K_1^- (s\bar{u})$  ($K_1^+(\bar{s}u)$) feels attraction (repulsion) in nuclear matter.
 This tendency is consistent with the expectation that nuclear matter attracts (repels) the $u$-anti-quark (the $u$-quark)
 as in the case of charged kaons, $K^- (s\bar{u})$ and $K^+(\bar{s}u)$~\cite{Martin:1980qe}.

\begin{figure}
\centerline{\includegraphics[width=10 cm]{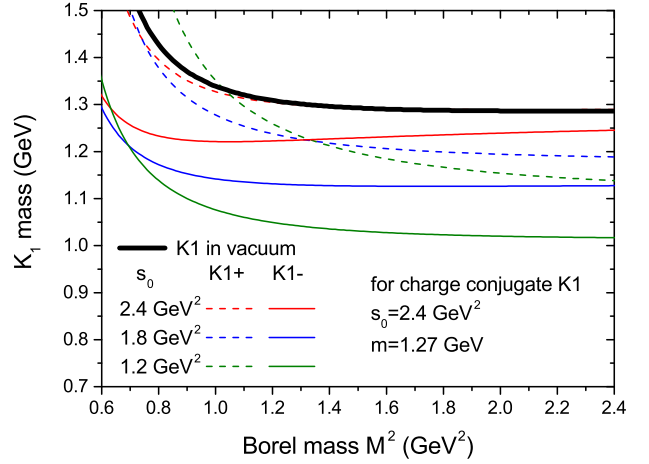}}
\caption{ The Borel curve for $K_1$ mass from $\Pi_1$ sum rule at nuclear matter density.  }
\label{Borel-mass-medium}
\end{figure}

The above results can be confirmed by using the traditional Borel stability analysis in nuclear matter as originally proposed in Ref.~\cite{Hatsuda:1991ez}.
Shown in  Figure \ref{Borel-mass-medium} are the Borel curves for the $K_1$ mass in the vacuum (the black curve) and those in the medium.
   The most stable Borel curve for $K_1^+$ occurs at $s_0=2.4$ GeV$^2$ with a slight decrease of the mass,
   while that for the $K_1^-$ occurs at a much smaller threshold with a large reduction of the mass consistent with the previous optimization method. Furthermore, one finds that for both of the charge states, the most stable curves have plateaux and extremums within the given Borel window.

\subsubsection{In-medium width}

So far, we have approximated the spectral function as the sum of a delta function for the ground state and a step function for excited states.
However, in-medium spectral function would have  more complicated structure, and the medium modification of the  OPE side is reflected  as a
combination of the mass and width changes in QCD sum rules (see e.g. the 3rd reference in ~\cite{Klingl:1997kf}).
In order to investigate this possibility, we replace  the delta function by the Breit-Wigner form,
\begin{eqnarray}
\delta(s-m_\pm^2) \rightarrow \frac{1}{\pi}\frac{\sqrt{s}~\Gamma_\pm}{(s-m_\pm^2)^2+s\Gamma_\pm^2},
\end{eqnarray}
 where $\Gamma_\pm$ is the width of $K_1^\pm$. We will allow the width to change by $\Gamma_{\pm} = \Gamma_{K_1}+\Gamma'_\pm \rho$ in  nuclear matter at low density.
We expand $F_{\pm}$,  $\Gamma_{\pm}$ and $s_0^{\pm} $ up to the linear order in $\rho$, while keeping $m_{\mp}(\rho)$ without expansion.  This is to probe the  maximum width change associated with the change in the OPE \cite{Leupold:1997dg}.
 Eq.~(\ref{eq:SR-m}) and Eq.~(\ref{minimum1}) are then modified as
\begin{eqnarray}
\mathcal{F}(M^2)F_\pm '+\mathcal{W}(M^2)\Gamma'_\pm +\mathcal{S}(M^2) {s'}_0^\pm =\mathcal{C}_\pm(M^2),\nonumber\\
\end{eqnarray}
where $\Gamma'_\pm $ is considered as variables instead of $m'_\pm $, and the terms proportional to $m'_\pm $ are moved into $\mathcal{C}_\pm (M^2)$ as
\begin{eqnarray}
\mathcal{F}(M^2)=-\frac{1}{\pi}\int_0^\infty \frac{s^{3/2}\Gamma_{K_1}}{(s-m_{K_1}^2)^2+s\Gamma_{K_1}^2}e^{-s/M^2}ds,~~~~~~&&\nonumber\\
\mathcal{W}(M^2)=-\frac{F_{K_1}}{\pi}\int_0^\infty \frac{s^{3/2}\Gamma_{K_1}}{(s-m_{K_1}^2)^2+s\Gamma_{K_1}^2}~~~~~~~~~~~~~~&&\nonumber\\
\times \bigg(1-\frac{2s\Gamma_{K_1}^2}{(s-m_{K_1}^2)^2+s\Gamma_{K_1}^2}\bigg)e^{-s/M^2}ds,&&\nonumber\\
\mathcal{S}(M^2)= S(M^2),~~~~~~~~~~~~~~~~~~~~~~~~~~~~~~~~~~~~~~~~~~~~~~&&\nonumber\\
 \mathcal{C}_\pm (M^2)=-m_s \langle\bar{u}u\rangle_N
+\frac{\alpha_s}{12\pi}\langle G^2\rangle_N ~~~~~~~~~~~~~~~~~~~~~~~~&&\nonumber\\
+\frac{m_N}{2}(A_2^u +A_2^s)\pm \frac{m_{K_1}}{3}(A_1^u-A_1^s)~~~~~~~~~~~~~~~&&\nonumber\\
+\frac{32\pi \alpha_s}{9 M^2}\bigg\{
\langle\bar{u}u\rangle_N \langle\bar{s}s\rangle_0 +\langle\bar{u}u\rangle_0 \langle\bar{s}s\rangle_N &&\nonumber\\
+\frac{2}{9}(\langle\bar{u}u\rangle_N\langle\bar{u}u\rangle_0+\langle\bar{s}s\rangle_N\langle\bar{s}s\rangle_0)\bigg\}&&\nonumber\\
-\frac{5}{6 M^2}m_N^3(A_4^u +A_4^s)\mp \frac{2m_{K_1} m_N^2}{3M^2}(A_3^u-A_3^s)~~~~~~&&\nonumber\\
+m'_\pm\bigg\{ \frac{F_{K_1}}{2\pi m_{K_1}}\int_0^\infty \frac{s^{3/2}\Gamma_{K_1}e^{-s/M^2}}{(s-m_{K_1}^2)^2+s\Gamma_{K_1}^2}~~~~~~~~~~~&&\nonumber\\
\times \bigg(1-\frac{8m_{K_1}^2(s-m_{K_1}^2)}{(s-m_{K_1}^2)^2+s\Gamma_{K_1}^2}\bigg)ds  \bigg\}~~~~~~~&&\nonumber\\
+m_\mp{'}\bigg\{  -\frac{F_{K_1}}{2\pi m_{K_1}}\int_0^\infty \frac{s^{3/2}\Gamma_{K_1}e^{-s/M^2}}{(s-m_{K_1}^2)^2+s\Gamma_{K_1}^2}ds \bigg\}~~&&\nonumber\\
+{s'}_0^\mp\bigg\{\frac{1}{2}\bigg(-1+\frac{m_{K_1}}{\sqrt{s_0}}\bigg)(B_0 s_0-B_2)e^{-s_0/M^2}\bigg\}.~~&& \nonumber
\end{eqnarray}

Defining the function $V(F'_\pm, \Gamma'_\pm, {s'}_0^\pm)$ similarly as before, the differential equations
\begin{eqnarray}
\frac{\partial V_\pm}{\partial F'_\pm}=\frac{\partial V_\pm}{\partial \Gamma'_\pm}=\frac{\partial V_\pm}{\partial {s'}_0^\pm}=0,\nonumber
\end{eqnarray}
are expressed as
\begin{widetext}
\begin{eqnarray}
F'_\pm \int dM^2 \mathcal{F}^2(M^2) +\Gamma'_\pm \int dM^2 \mathcal{F}(M^2)\mathcal{M}(M^2)
+{s'}_0^\pm\int
dM^2 \mathcal{F}(M^2)\mathcal{S}(M^2)&=& \int dM^2 \mathcal{F}(M^2) \mathcal{C}_\pm(M^2),\nonumber\\
F'_\pm \int dM^2 \mathcal{F}(M^2)\mathcal{M}(M^2) +\Gamma'_\pm \int dM^2 \mathcal{M}^2(M^2)
+{s'}_0^\pm \int
dM^2 M(M^2)S(M^2)&=& \int dM^2 M(M^2) \mathcal{C}_\pm(M^2),\nonumber\\
F'_\pm \int dM^2 \mathcal{F}(M^2)\mathcal{S}(M^2) +\Gamma'_\pm \int dM^2 \mathcal{M}(M^2)\mathcal{S}(M^2)
+{s'}_0^\pm \int dM^2 \mathcal{S}^2(M^2)&=& \int dM^2 \mathcal{S}(M^2) \mathcal{C}_\pm(M^2).
\label{eq:sum-w}
\end{eqnarray}
\end{widetext}

\begin{figure}[t]
\centerline{
\includegraphics[width=9.0 cm]{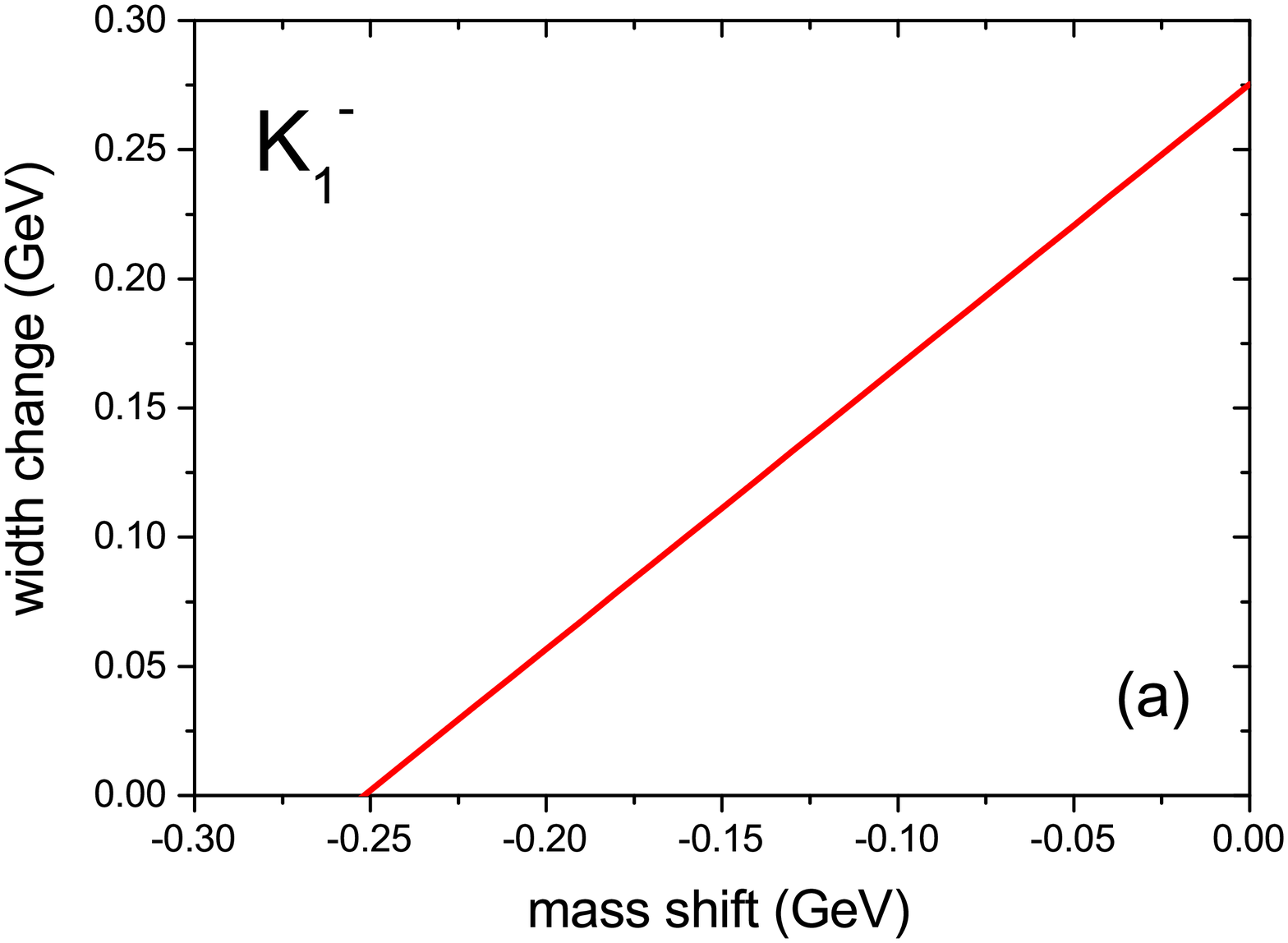}}
\centerline{
\includegraphics[width=9.0 cm]{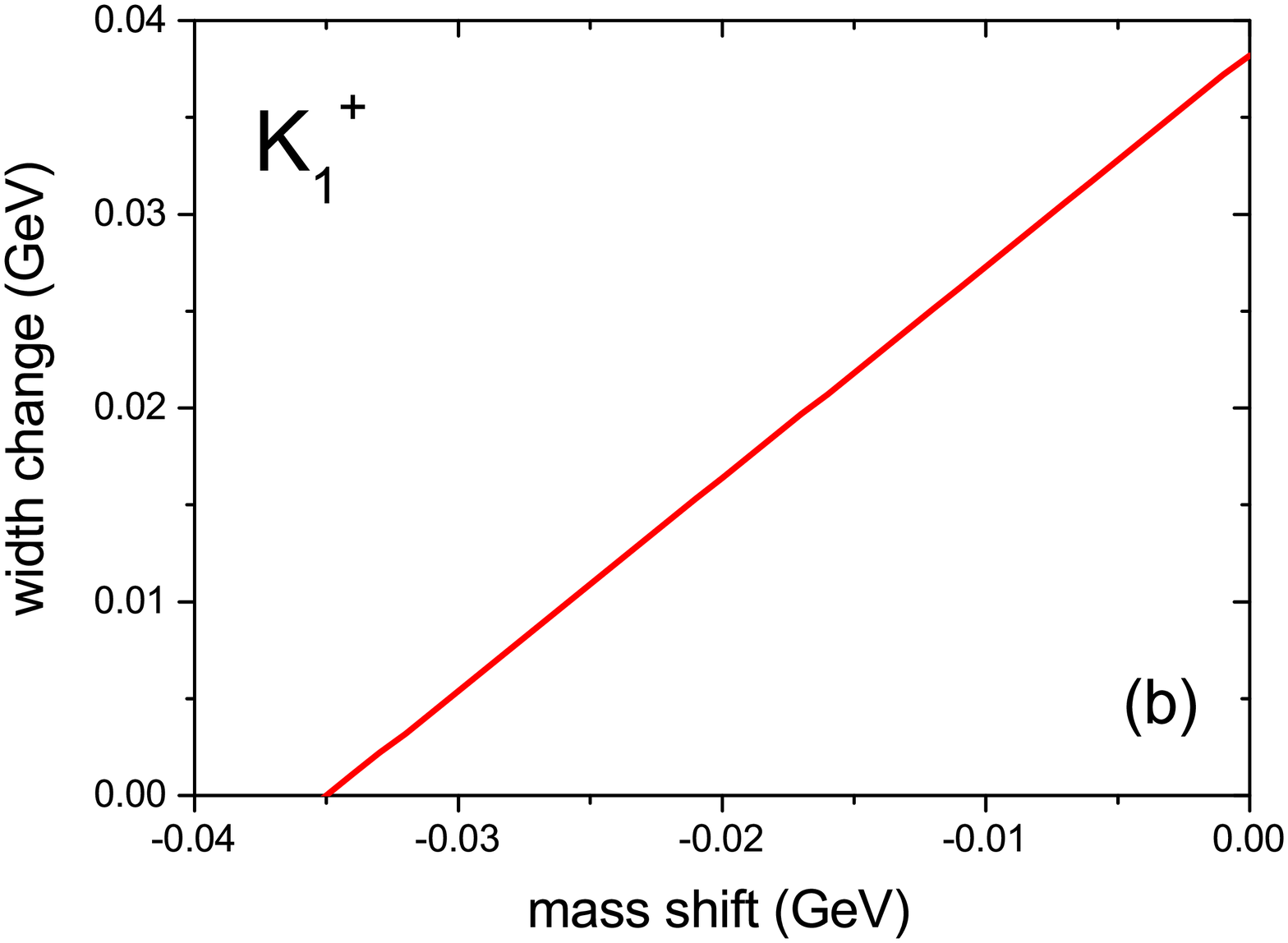}}
\caption{The constraints on the mass shift and the width change of $K_1^\pm$ in the nuclear matter}
\label{k1-medium}
\end{figure}

We find that Eq.~(\ref{eq:sum-w}) has only a weak dependence on the vacuum width $\Gamma_{K_1}$, so that we take
$\Gamma_{K_1}=0$ in solving the coupled equations  since the input parameters $F_{K_1}$ and $s_0$ were  obtained in this limit.
Figure~\ref{k1-medium} shows the constraints on the mass modification, $\delta m_{\pm} (\rho) = m_{\pm} (\rho) - m_{K_1}$
 and the width modification $\delta \Gamma_{\pm} (\rho) = \Gamma'_\pm\rho$ obtained from Eq.~(\ref{eq:sum-w}).
 As for $K_1^-$, the maximum change of the width is +275 MeV, while +38 MeV for $K_1^+$.
The decay of the $K_1 $ is dominated by $K  \rho$ ($42\pm 6$\%) in vacuum.
 Therefore, keeping the $K_1$ mass the same, if the mass of $K^-$ ($K^+$)
decreases (remains the same) in the medium, the phase space for the
corresponding   $K \rho$ decay for $K_1^-$ ($K_1^+$) will increase (remain
the same), which provides a possible physical mechanism for their
asymmetric width change in medium.

\section{Summary and discussions}

In this paper, we have carried out QCD sum rule analyses for the  open strange meson $K_1$ in the vacuum and in the medium.
We first show that  the experimental mass of $K_1 (1270)$ in the vacuum  is well reproduced by the QCD condensates with  a proper  choice of the
  continuum threshold and the Borel window.  In nuclear matter,    $K_1^{-} $ and $K_1^{+} $ become non-degenerate  due to the breaking of the charge conjugation invariance.
 By extracting the ground state of each charge state from the polarization functions and by formulating coupled QCD sum rules in nuclear matter,
 we  obtained a relation between the in-medium mass and width of $K_1^\pm$  through the density dependence of the  scalar and twist-two condensates.
 In particular, the upper limit of the mass shifts  for $K_1^-$ and $K_1^+$  without width modifications are -249 MeV and -35 MeV, respectively,
 which indicates that   $K_1^- (s\bar{u})$   ($K_1^- (\bar{s}u)$) feels attraction (repulsion) in nuclear matter.
 Once  the change of the widths is allowed, however, those mass shifts get smaller.
Furthermore, the $K_1(1270)$ and $K_1(1400)$ are the analogues of $a_1$ and $b_1$ mesons and can be analyzed through the axial vector as well as tensor currents \cite{Jeong:2018exh}.  A more detailed discussion on how the these currents couple to $K_1(1270)$ and $K_1(1400)$  in the medium together with their respective mass changes are an important topic for future investigations.

Experimentally $K_1^-$ is known to be  produced through  $K^-$-nucleon reactions \cite{Gavillet:1978rj,Daum:1981hb},
so that   the modification of $K_1^-$ in nuclei would be best searched through the $K^-$ reaction on various nuclear targets.
 Such experimental possibility may be provided by the  kaon beam at J-PARC with the energy up to 2.0 GeV.
 Maximum $\sqrt{s}$ of a kaon and a nucleon is  2.23 GeV ignoring the fermi motion, and is as large as 2.48 GeV  including the fermi motion.
 Those numbers are close to the threshold value of the $K_1^-$ production which is around 2.2 GeV.
  The measurement of hadronic decays ($K_1 \rightarrow K~\rho$ and $K_1 \rightarrow K^* \pi$)
  as well as the measurement of the  excitation function  would be possible probes to detect  the spectral  shift of  $K_1^-$.
  Furthermore,  similar experiments for the $K^*$ (the chiral partner of $K_1$) will give model-independent
  estimate of the  chiral order parameter in the medium, and hence provide crucial hints to the  partial restoration of chiral symmetry  in nuclear medium.

\section*{Acknowledgements}
This work was supported by the Korea National Research Foundation (NRF) under Grant No. 2016R1D1A1B03930089.
T.H. was partially supported by RIKEN iTHEMS program and JSPS Grant-in-Aid for Scientific Research (S),  No.18H05236.
The work of T.H. was performed at the Aspen Center for Physics, which is supported by National Science Foundation grant PHY-1607611.


\hfil\break
\centerline{\bf \large Appendix A}
\bigskip

Here we show that the multiplicative factor to the step function for the continuum part corresponds to the perturbative part of OPE side.
Suppose
\begin{eqnarray}
\frac{1}{\pi}{\rm Im}\Pi_1 (s)=B_0 s-B_2+{\rm pole ~contribution},
\end{eqnarray}
\smallskip
where the first two terms on the right hand side represent the continuum contribution.
Taking the Borel transformation to the dispersion relation, we find

\begin{eqnarray}
&&\hat{B}\frac{1}{\pi}\int \frac{{\rm Im} \Pi_1 (s)}{s+Q^2}ds \nonumber\\
&&=\frac{1}{M^2}\int^\infty_{0^+} \bigg(B_0 s- B_2 \bigg)e^{-s/M^2}ds +{\rm pole ~contribution}\nonumber\\
\nonumber\\
&&= B_0 M^2 -B_2+{\rm pole ~contribution}.
\label{continuum2}
\end{eqnarray}
We find that the first two terms are exactly the same as the perturbative part of the OPE in the dispersion relation.

\hfil\break
\appendix
\centerline{\bf \large Appendix B}
\bigskip

Using the integral form of step function
\begin{eqnarray}
\theta(t)=\frac{1}{2\pi i}\int_{-\infty}^{\infty}dw' \frac{e^{iw't}}{w'-i\varepsilon}, \nonumber
\end{eqnarray}
together with the decomposition of identity operator
\begin{eqnarray}
I \equiv \sum_\lambda \int \frac{d^3 {\textbf p}}{(2\pi)^3}\frac{1}{2w_{_{\textbf p}}}|\lambda_{_{\textbf p}} \rangle\langle\lambda_{_{\textbf p}}| \nonumber
\end{eqnarray}
 with  the state ($|\lambda_{_{\textbf p}} \rangle$)
 having definite polarization and dispersion relation $w_{_{\textbf p}} =\sqrt{m_\lambda^2+{\textbf p}^2}$,
 the time ordered correlation function reads \cite{Morath:1999cv}
\begin{eqnarray}
&&\Pi_{\mu \nu}(q)= i\int d^4 x \ e^{iq\cdot x}\langle 0| T[\bar{u}\gamma_\mu \gamma_5
s(x), \bar{s}\gamma_\nu \gamma_5 u(0)]| 0\rangle\nonumber\\
&&=\sum_\lambda \int d^4 x \ e^{iq\cdot x}\int \frac{dw'}{2\pi}
\frac{1}{w'-i\varepsilon} \int \frac{d^3 {\textbf p}}{(2\pi)^3}\frac{1}{2w_{_{\textbf p}}}\nonumber\\
&&\times \bigg\{ e^{iw't} \langle 0|\bar{u}\gamma_\mu \gamma_5
s(x) |\lambda_{_{\textbf p}} \rangle\langle\lambda_{_{\textbf p}}|\bar{s}\gamma_\nu \gamma_5 u(0)|0\rangle\nonumber\\
&&~~+e^{-iw't} \langle 0|\bar{s}\gamma_\nu \gamma_5 u(0)|\lambda_{_{\textbf p}} \rangle\langle\lambda_{_{\textbf p}}|\bar{u}\gamma_\mu \gamma_5
s(x)|0\rangle\bigg\}.\nonumber\\
\end{eqnarray}

Using the translation operators,
\begin{eqnarray}
\langle 0|\bar{u}\gamma_\mu \gamma_5 s(x) |\lambda_{_{\textbf p}} \rangle=\langle 0|\bar{u}\gamma_\mu \gamma_5 s(0) |\lambda_{_{\textbf p}} \rangle e^{-i(w_{\bf p}t-{\bf p\cdot x})},\nonumber\\
\langle\lambda_{_{\textbf p}}|\bar{u}\gamma_\mu \gamma_5 s(x)|0\rangle=\langle\lambda_{_{\textbf p}}|\bar{u}\gamma_\mu \gamma_5
s(0)|0\rangle e^{i(w_{\bf p}t-{\bf p\cdot x})},\nonumber\\
\end{eqnarray}

the polarization function is expressed as
\begin{widetext}
\begin{eqnarray}
\Pi_{\mu \nu}(q)&=&\sum_\lambda  \int \frac{dw'}{w'-i\varepsilon} \bigg\{ \frac{1}{2w_{_{\textbf q}}}\delta (q_0-w_{_{\textbf q}}+w') \langle 0|\bar{u}\gamma_\mu \gamma_5
s(0) |\lambda_{_{\textbf q}} \rangle\langle\lambda_{_{\textbf q}}|\bar{s}\gamma_\nu \gamma_5 u(0)|0\rangle \nonumber\\
&&~~~~~~~~~~~~~~~+\frac{1}{2w_{-{\textbf q}}}\delta (q_0+w_{-{\textbf q}}-w') \langle 0|\bar{s}\gamma_\nu \gamma_5 u(0)|\lambda_{-{\textbf q} } \rangle\langle\lambda_{-{\textbf q}}|\bar{u}\gamma_\mu \gamma_5
s(x)|0\rangle\bigg\}\nonumber\\
&=&\sum_\lambda  \bigg\{\frac{1}{2w_{_{\textbf q}}(w_{_{\textbf q}}-q_0-i\varepsilon)}\langle 0|\bar{u}\gamma_\mu \gamma_5
s(0) |\lambda_{_{\textbf q}} \rangle\langle\lambda_{_{\textbf q}}|\bar{s}\gamma_\nu \gamma_5 u(0)|0\rangle \nonumber\\
&&~~~+\frac{1}{2w_{-{\textbf q}}(w_{-{\textbf q}}+q_0-i\varepsilon)} \langle 0|\bar{s}\gamma_\nu \gamma_5 u(0)|\lambda_{-{\textbf q} } \rangle\langle\lambda_{-{\textbf q}}|\bar{u}\gamma_\mu \gamma_5
s(x)|0\rangle\bigg\}.
\end{eqnarray}
Furthermore, using
\begin{eqnarray}
\frac{1}{w_{_{\textbf q}}-q_0-i\varepsilon}=\frac{w_{_{\textbf q}}+q_0+i\varepsilon}{w_{_{\textbf q}}^2-q_0^2-2i\varepsilon w_{_{\textbf q}}}
=\frac{P}{w_{_{\textbf q}}-q_0}+i\pi(w_{_{\textbf q}}+q_0) \delta(w_{_{\textbf q}}^2-q_0^2),
\end{eqnarray}
where $\varepsilon w_{_{\textbf q}}$ is definitely positive, and the same for the second term, the imaginary part of correlation function reads
\begin{eqnarray}
\frac{1}{\pi} {\rm Im} \Pi_{\mu \nu}(q)&=&\sum_\lambda  \bigg\{\bigg(\frac{1}{2}+\frac{q_0}{2w_{_{\textbf q}}}\bigg)\langle 0|\bar{u}\gamma_\mu \gamma_5
s(0) |\lambda_{_{\textbf q}} \rangle\langle\lambda_{_{\textbf q}}|\bar{s}\gamma_\nu \gamma_5 u(0)|0\rangle\delta(q_0^2-w_{_{\textbf q}}^2)\nonumber\\
&&~~~+\bigg(\frac{1}{2}-\frac{q_0}{2w_{-{\textbf q}}}\bigg) \langle 0|\bar{s}\gamma_\nu \gamma_5 u(0)|\lambda_{-{\textbf q} } \rangle\langle\lambda_{-{\textbf q}}|\bar{u}\gamma_\mu \gamma_5
s(x)|0\rangle\delta(q_0^2-w_{-{\textbf q}}^2)\bigg\}.
\label{imaginary}
\end{eqnarray}
\end{widetext}

Note that  $|\lambda_{_{\textbf p}} \rangle$ can be any state whose quantum number is the same as that of $\bar{u}\gamma_\mu \gamma_5 s$.
This interpolating field is coupled to both the pseudoscalar $K$ meson and the axial vector $K_1$ meson:

\begin{eqnarray}
&&\langle 0|\bar{u}\gamma_\mu \gamma_5 s |K^- \rangle=\langle 0|\bar{s}\gamma_\mu \gamma_5 u |K^+ \rangle=if_Kq_\mu \nonumber\\
&&\langle 0|\bar{u}\gamma_\mu \gamma_5 s |K_1^-\rangle=\langle 0|\bar{s}\gamma_\mu \gamma_5 u |K_1^+\rangle=\frac{m_{K_1}^2}{g_{K_1}}\epsilon_\mu,\nonumber\\
\label{field-strength}
\end{eqnarray}
where kaon decay constant $f_K=160$ MeV, and $g_{K_1}$, $\epsilon_\mu$ are the coupling constant and polarization vector of $K_1$ respectively. The operator $\bar{u}\gamma_\mu \gamma_5 s$ couples to $K_1^-$ state, because it has the annihilation operator of $K_1^-$, and $\bar{s}\gamma_\mu \gamma_\nu u$ couples to $K_1^+$ for the same reason. Furthermore, the overlap strength to their respective states for both fields  are the same in vacuum.
Substituting Eq.~(\ref{field-strength}) into Eq.~(\ref{imaginary}),

\begin{eqnarray}
&&\frac{1}{\pi} {\rm Im} \Pi_{\mu \nu}(q^2)= q_\mu q_\nu f_K^2 \delta(q^2-m_K^2)\nonumber\\
&&~~~~~~~~+\bigg(-g_{\mu \nu}+\frac{q_\mu q_\nu }{q^2}\bigg)\frac{m_{K_1}^4}{g_{K_1}^2}\delta(q^2-m_{K_1}^2)\nonumber\\
&&~~~~~~+\sum_{K_1^*} \bigg(-g_{\mu \nu}+\frac{q_\mu q_\nu }{q^2}\bigg)\frac{m_{K_1^*}^4}{g_{K_1^*}^2}\delta(q^2-m_{K_1^*}^2),
\label{imaginary2}
\end{eqnarray}
where $K_1^*$ represents excited state.
Decomposing Eq.~(\ref{imaginary2}) into $\Pi_1$ and $\Pi_2$,

\begin{eqnarray}
\frac{1}{\pi}{\rm Im}\Pi_1(q^2)&=&\frac{m_{K_1}^4}{g_{K_1}^2}\delta(q^2-m_{K_1}^2)+\sum_{K_1^*} \frac{m_{K_1^*}^4}{g_{K_1^*}^2}\delta(q^2-m_{K_1^*}^2),\nonumber\\
\frac{1}{\pi}{\rm Im}\Pi_2(q^2)&=&f_K^2 \delta(q^2-m_K^2)+\frac{m_{K_1}^2}{g_{K_1}^2}\delta(q^2-m_{K_1}^2)\nonumber\\
&&~~+\sum_{K_1^*} \frac{m_{K_1^*}^2}{g_{K_1^*}^2}\delta(q^2-m_{K_1^*}^2).
\end{eqnarray}

Since the excited states have broad widths and overlap with each other, they can be simplified into a step function with a threshold value $s_0$,

\begin{eqnarray}
\frac{1}{\pi}{\rm Im}\Pi_1 (q^2)&=&  \frac{m_{K_1}^4}{g_{K_1}^2}\delta(q^2-m_{K_1}^2)\nonumber\\
&&+(B_0 q^2-B_2) \theta(q^2-s_0),\label{pi1-app}\\
\frac{1}{\pi}{\rm Im}\Pi_2(q^2)&=&f_K^2 \delta(q^2-m_K^2)+\frac{m_{K_1}^2}{g_{K_1}^2}\delta(q^2-m_{K_1}^2)\nonumber\\
&&+ B_0 \theta(q^2-s_0).
\end{eqnarray}
The multiplicative factors of the step functions correspond to the perturbative part of OPE side, as shown in Appendix A.

Since the charge conjugation between $K_1^-$ and $K_1^+$ is broken in nuclear matter and
the Lorentz invariance is broken in the rest frame of nuclear matter, Eq. (\ref{field-strength}) changes into
\begin{eqnarray}
\langle {\rm n.m.}|\bar{u}\gamma_\mu \gamma_5 s |{\rm n.m.}+K_1^-\rangle&=&\frac{m_{K_1^-}^2}{g_{K_1^-}}\epsilon_\mu,\nonumber\\
\langle {\rm n.m.}|\bar{s}\gamma_\mu \gamma_5 u |{\rm n.m.}+K_1^+\rangle&=&\frac{m_{K_1^+}^2}{g_{K_1^+}}\epsilon_\mu.
\end{eqnarray}
in the pole approximation at ${\textbf q} =0$.  Here $ | {\rm n.m.} \rangle $ implies the ground state of nuclear matter.
 Then Eq. (\ref{pi1-app}) at ${\textbf q} =0$   changes into
\begin{eqnarray}
\frac{1}{\pi}{\rm Im}\Pi_1(q^2)&=&\bigg(\frac{1}{2}+\frac{q_0}{2m_{K_1^-}}\bigg)\frac{m_{K_1^-}^4}{g_{K_1^-}^2}\delta(q^2-m_{K_1^-}^2)\nonumber\\
&+&\bigg(\frac{1}{2}-\frac{q_0}{2m_{K_1^+}}\bigg)\frac{m_{K_1^+}^4}{g_{K_1^+}^2}\delta(q^2-m_{K_1^+}^2)\nonumber\\
&+&\sum_{K_1^{-*}}\bigg(\frac{1}{2}+\frac{q_0}{2m_{K_1^{-*}}}\bigg) \frac{m_{K_1^{-*}}^4}{g_{K_1^{-*}}^2}\delta(q^2-m_{K_1^{-*}}^2)\nonumber\\
&+&\sum_{K_1^{+*}}\bigg(\frac{1}{2}-\frac{q_0}{2m_{K_1^{+*}}}\bigg) \frac{m_{K_1^{+*}}^4}{g_{K_1^{+*}}^2}\delta(q^2-m_{K_1^{+*}}^2),\nonumber\\
\end{eqnarray}
which is decomposed into even and odd dimensions~\cite{Kondo:1994uw,Jido:1996ia}:

\begin{eqnarray}
\frac{1}{\pi}{\rm Im}\Pi_1=\frac{1}{\pi}({\rm Im}\Pi^e +q_0 {\rm Im}\Pi^o),
\end{eqnarray}
where
\begin{eqnarray}
&&\frac{1}{\pi}{\rm Im} \Pi^e=\frac{m_{K_1^-}^4}{2 g_{K_1^-}^2}\delta(q^2-m_{K_1^-}^2)+\frac{m_{K_1^+}^4}{2 g_{K_1^+}^2}\delta(q^2-m_{K_1^+}^2)\nonumber\\
&&+\sum_{K_1^{-*}}\frac{m_{K_1^{-*}}^4}{2g_{K_1^{-*}}^2}\delta(q^2-m_{K_1^{-*}}^2)
+\sum_{K_1^{+*}}\frac{m_{K_1^{+*}}^4}{2g_{K_1^{+*}}^2}\delta(q^2-m_{K_1^{+*}}^2),\nonumber\\
&&\frac{1}{\pi}{\rm Im}\Pi^o=\frac{m_{K_1^-}^3}{2g_{K_1^-}^2}\delta(q^2-m_{K_1^-}^2)-\frac{m_{K_1^+}^3}{2g_{K_1^+}^2}\delta(q^2-m_{K_1^+}^2)\nonumber\\
&&+\sum_{K_1^{-*}}\frac{m_{K_1^{-*}}^3}{2g_{K_1^{-*}}^2}\delta(q^2-m_{K_1^{-*}}^2)
-\sum_{K_1^{+*}}\frac{m_{K_1^{+*}}^3}{2g_{K_1^{+*}}^2}\delta(q^2-m_{K_1^{+*}}^2).\nonumber
\end{eqnarray}

$K_1^\pm$ poles are separated from the following linear combinations of $\Pi^e$ and $\Pi^o$

\begin{eqnarray}
&&\frac{2}{\pi}{\rm Im}(\Pi^e +m_{K_1^+} \Pi^o)= \frac{m_{K_1^-}^4}{ g_{K_1^-}^2}
\bigg(1+\frac{m_{K_1^+}}{m_{K_1^-}}\bigg)\delta(q^2-m_{K_1^-}^2)\nonumber\\
&&~~~~~~~~~~ +(B_0 q^2-B_2)\bigg\{ (1+\frac{m_{K_1^+}}{q_0})\theta(q^2-s_0^-)\nonumber\\
&&~~~~~~~~~~~~~~~~~~~~~~~~~~~~~+(1-\frac{m_{K_1^+}}{q_0})\theta(q^2-s_0^+)\bigg\}\nonumber\\
&&\frac{2}{\pi}{\rm Im}(\Pi^e -m_{K_1^-} \Pi^o)= \frac{m_{K_1^+}^4}{ g_{K_1^+}^2}
\bigg(1+\frac{m_{K_1^-}}{m_{K_1^+}}\bigg)\delta(q^2-m_{K_1^+}^2)\nonumber\\
&&~~~~~~~~~~ +(B_0 q^2-B_2)\bigg\{ (1+\frac{m_{K_1^-}}{q_0})\theta(q^2-s_0^+)\nonumber\\
&&~~~~~~~~~~~~~~~~~~~~~~~~~~~~~+(1-\frac{m_{K_1^-}}{q_0})\theta(q^2-s_0^-)\bigg\},
\end{eqnarray}
where the continuum parts are replaced by the step functions:

\begin{eqnarray}
&&\sum_{K_1^{\pm *}}\frac{m_{K_1^{\pm *}}^4}{g_{K_1^{\pm *}}^2}\delta(q^2-m_{K_1^{\pm *}}^2)\rightarrow\nonumber\\
&& ~~~~~~~~~~~~~~ (B_0 q^2-B_2)\theta(q^2-s_0^\pm)\nonumber
\end{eqnarray}
and
\begin{eqnarray}
&&\sum_{K_1^{\pm *}}\frac{m_{K_1^{\pm *}}^3}{g_{K_1^{\pm *}}^2}\delta(q^2-m_{K_1^{\pm *}}^2)\rightarrow\nonumber\\
&& ~~~~~~~~~~~~~~ (B_0 q^2-B_2)\frac{1}{q_0}\theta(q^2-s_0^\pm).
\end{eqnarray}




\begin{thebibliography}{10}

\bibitem{Weinberg:1967kj}
  S.~Weinberg,
  Phys.\ Rev.\ Lett.\  {\bf 18}, 507 (1967).

\bibitem{Shifman:1978bx}
  M.~A.~Shifman, A.~I.~Vainshtein and V.~I.~Zakharov,
  Nucl.\ Phys.\  B {\bf 147}, 385 (1979); 
  M.~A.~Shifman, A.~I.~Vainshtein and V.~I.~Zakharov,
  Nucl.\ Phys.\  B {\bf 147}, 448 (1979).


\bibitem{Hatsuda:1985eb}
  T.~Hatsuda and T.~Kunihiro,
  Phys.\ Rev.\ Lett.\  {\bf 55}, 158 (1985).

\bibitem{Brown:1991kk}
  G.~E.~Brown and M.~Rho,
  Phys.\ Rev.\ Lett.\  {\bf 66}, 2720 (1991).

\bibitem{Hatsuda:1991ez}
  T.~Hatsuda and S.~H.~Lee,
  Phys.\ Rev.\ C {\bf 46}, no. 1, R34 (1992).

\bibitem{Klingl:1997kf}
  F.~Klingl, N.~Kaiser and W.~Weise,
  Nucl.\ Phys.\ A {\bf 624}, 527 (1997).
%
  R.~Rapp and J.~Wambach,
  Adv.\ Nucl.\ Phys.\  {\bf 25}, 1 (2000).
%
  S.~Leupold, V.~Metag and U.~Mosel,
  Int.\ J.\ Mod.\ Phys.\ E {\bf 19}, 147 (2010).
%
  P.~Gubler and W.~Weise,
  Nucl.\ Phys.\ A {\bf 954}, 125 (2016).


\bibitem{Suzuki:2002ae}
  K.~Suzuki {\it et al.},
  Phys.\ Rev.\ Lett.\  {\bf 92}, 072302 (2004).
%
  P.~Kienle and T.~Yamazaki,
  Prog.\ Part.\ Nucl.\ Phys.\  {\bf 52}, 85 (2004).

\bibitem{Kolomeitsev:2002gc}
  E.~E.~Kolomeitsev, N.~Kaiser and W.~Weise,
  Phys.\ Rev.\ Lett.\  {\bf 90}, 092501 (2003).
%
  D.~Jido, T.~Hatsuda and T.~Kunihiro,
  Phys.\ Lett.\ B {\bf 670}, 109 (2008).

\bibitem{Hayano:2008vn}
 R.~S.~Hayano and T.~Hatsuda,
  Rev.\ Mod.\ Phys.\  {\bf 82}, 2949 (2010).



\bibitem{Naruki:2005kd}
  M.~Naruki {\it et al.},
  Phys.\ Rev.\ Lett.\  {\bf 96}, 092301 (2006).


\bibitem{Muto:2005za}
  R.~Muto {\it et al.}  [KEK-PS-E325 Collaboration],
  Phys.\ Rev.\ Lett.\  {\bf 98}, 042501 (2007).

\bibitem{Ichikawa:2018woh}
  M.~Ichikawa {\it et al.},
  arXiv:1806.10671 [physics.ins-det].

\bibitem{Trnka:2005ey}
  D.~Trnka {\it et al.}  [CBELSA/TAPS Collaboration],
  Phys.\ Rev.\ Lett.\  {\bf 94}, 192303 (2005).

\bibitem{Gubler:2016djf}
  P.~Gubler, T.~Kunihiro and S.~H.~Lee,
  Phys.\ Lett.\ B {\bf 767}, 336 (2017).

\bibitem{DuttMazumder:2000ys}
  A.~K.~Dutt-Mazumder, R.~Hofmann and M.~Pospelov,
  Phys.\ Rev.\ C {\bf 63}, 015204 (2001).


\bibitem{Thomas:2005dc}
  R.~Thomas, S.~Zschocke and B.~Kampfer,
  Phys.\ Rev.\ Lett.\  {\bf 95}, 232301 (2005).


\bibitem{Leupold:2009kz}
  S.~Leupold, V.~Metag and U.~Mosel,
  Int.\ J.\ Mod.\ Phys.\ E {\bf 19}, 147 (2010).

\bibitem{Metag:2017yuh}
  V.~Metag, M.~Nanova and E.~Y.~Paryev,
  Prog.\ Part.\ Nucl.\ Phys.\  {\bf 97}, 199 (2017).



\bibitem{Dickson:2016gwc}
  R.~Dickson {\it et al.} [CLAS Collaboration],
  Phys.\ Rev.\ C {\bf 93}, 065202 (2016).

\bibitem{Lee:2013es}
  S.~H.~Lee and S.~Cho,
  Int.\ J.\ Mod.\ Phys.\ E {\bf 22}, 1330008 (2013).




\bibitem{Suzuki:1993yc}
  M.~Suzuki,
  Phys.\ Rev.\ D {\bf 47}, 1252 (1993).

\bibitem{Hatsuda:1997ev}
  T.~Hatsuda,
  arXiv:nucl-th/9702002.

%
%

\bibitem{Jido:1996ia}
  D.~Jido, N.~Kodama and M.~Oka,
  Phys.\ Rev.\  D {\bf 54}, 4532 (1996).


\bibitem{Suzuki:2015est}
  K.~Suzuki, P.~Gubler and M.~Oka,
  Phys.\ Rev.\ C {\bf 93}, no. 4, 045209 (2016).


\bibitem{Kondo:2005ur}
  Y.~Kondo, O.~Morimatsu and T.~Nishikawa,
  Nucl.\ Phys.\ A {\bf 764}, 303 (2006).



\bibitem{Furnstahl:1992pi}
  R.~J.~Furnstahl, D.~K.~Griegel and T.~D.~Cohen,
  Phys.\ Rev.\  C {\bf 46}, 1507 (1992).


\bibitem{Tanabashi:2018oca}
  M.~Tanabashi {\it et al.} [Particle Data Group],
  Phys.\ Rev.\ D {\bf 98}, no. 3, 030001 (2018).

\bibitem{Dominguez:2014pga}
  C.~A.~Dominguez, L.~A.~Hernandez and K.~Schilcher,
  JHEP {\bf 1507}, 110 (2015).

\bibitem{Novikov:1984rf}
  V.~A.~Novikov, M.~A.~Shifman, A.~I.~Vainshtein and V.~I.~Zakharov,
  Nucl.\ Phys.\ B {\bf 249}, 445 (1985)
  [Yad.\ Fiz.\  {\bf 41}, 1063 (1985)].

\bibitem{Lee:1989qj}
  S.~H.~Lee,
  Phys.\ Rev.\ D {\bf 40}, 2484 (1989).

\bibitem{Hilger:2011cq}
  T.~Hilger, B.~Kampfer and S.~Leupold,
  Phys.\ Rev.\ C {\bf 84}, 045202 (2011).

\bibitem{Buchheim:2014rpa}
  T.~Buchheim, T.~Hilger and B.~Kampfer,
  Phys.\ Rev.\ C {\bf 91}, 015205 (2015).


\bibitem{Reinders:1981ww}
  L.~J.~Reinders, S.~Yazaki and H.~R.~Rubinstein,
  Nucl.\ Phys.\ B {\bf 196}, 125 (1982).



\bibitem{Friman:1999wu}
  B.~Friman, S.~H.~Lee and H.~C.~Kim,
  Nucl.\ Phys.\ A {\bf 653}, 91 (1999).


\bibitem{Martin:2009iq}
  A.~D.~Martin, W.~J.~Stirling, R.~S.~Thorne and G.~Watt,
  Eur.\ Phys.\ J.\ C {\bf 63}, 189 (2009).



\bibitem{Martin:1980qe}
  A.~D.~Martin,
  Nucl.\ Phys.\  B {\bf 179}, 33 (1981).
  J.~S.~Hyslop, R.~A.~Arndt, L.~D.~Roper and R.~L.~Workman,
  Phys.\ Rev.\  D {\bf 46}, 961 (1992).
L.~Tolos, C.~Garcia-Recio, R.~Molina, J.~Nieves, E.~Oset, A.~Ramos and L.~L.~Salcedo,
  PoS ConfinementX {\bf }, 015 (2012).


\bibitem{Leupold:1997dg}
  S.~Leupold, W.~Peters and U.~Mosel,
  Nucl.\ Phys.\  A {\bf 628}, 311 (1998).

\bibitem{Jeong:2018exh}
  K.~S.~Jeong, S.~H.~Lee and Y.~Oh,
  JHEP {\bf 1808}, 179 (2018).

\bibitem{Gavillet:1978rj}
  P.~Gavillet {\it et al.}  [Amsterdam-CERN-Nijmegen-Oxford Collaboration],
  Phys.\ Lett.\  B {\bf 76}, 517 (1978).


\bibitem{Daum:1981hb}
  C.~Daum {\it et al.}  [ACCMOR Collaboration],
  Nucl.\ Phys.\  B {\bf 187}, 1 (1981).


\bibitem{Morath:1999cv}
  P.~Morath, W.~Weise and S.~H.~Lee,
{\it Prepared for 17th Autumn School: QCD: Perturbative or Nonperturbative? (AUTUMN 99), Lisbon, Portugal, 29 Sep - 4 Oct 1999}; Ph. D. thesis of P.~Morath 2001 Schwere Quarks in dichter Materie, Technische Universit\"{a}t M\"{u}nchen.

\bibitem{Kondo:1994uw}
  Y.~Kondo, O.~Morimatsu and Y.~Nishino,
  Phys.\ Rev.\  C {\bf 53}, 1927 (1996).






\end{thebibliography}
\end{document}